\documentclass[traditabstract,print]{aa}
\usepackage{graphicx}
\usepackage{epsfig}
\usepackage{amssymb,amsmath}
\usepackage{natbib}
\bibpunct{(}{)}{;}{a}{}{,}
\usepackage{pslatex}
\usepackage{caption}
\usepackage{longtable}
\usepackage{hyperref}

\begin{document}

\title{Measuring dark energy with the $E_{\rm iso}-E_{\rm p}$ correlation \\
of gamma-ray bursts using model-independent methods}

\author{J. S. Wang$^{1}$, F. Y. Wang$^{1,2,3}$
\thanks{fayinwang@nju.edu.cn(FYW)}, K. S. Cheng$^{2}$, and Z. G. Dai$^{1,3}$}

\institute{
$^{1}$ School of Astronomy and Space Science, Nanjing University, Nanjing 210093, China\\
$^2$ Department of Physics, University of Hong Kong, Pokfulam Road, Hong Kong, China\\
$^{3}$ Key Laboratory of Modern Astronomy and Astrophysics (Nanjing University),
Ministry of Education, Nanjing 210093, China }


\authorrunning{Wang et al.}
\titlerunning{Measuring dark energy with $E_{\rm iso}-E_{\rm p}$ correlation}


\abstract
{We use two model-independent methods to standardize
long gamma-ray bursts (GRBs) using the $E_{\rm iso}-E_{\rm p}$
correlation ($\log E_{\rm iso}=a+b\log E_{\rm p}$), where $E_{\rm
iso}$ is the isotropic-equivalent gamma-ray energy and $E_{\rm p}$
is the spectral peak energy. We update 42 long GRBs and attempt
to  constrain the cosmological parameters. The full sample
contains 151 long GRBs with redshifts from 0.0331 to 8.2. The first
method is the simultaneous fitting method. We take the extrinsic scatter
$\sigma_{\rm ext}$   into account and assign it to the
parameter $E_{\rm iso}$. The best-fitting values are
$a=49.15\pm0.26$, $b=1.42\pm0.11$, $\sigma_{\rm ext}=0.34\pm0.03$
and $\Omega_m=0.79$ in the flat $\Lambda$CDM model. The constraint
on $\Omega_m$ is $0.55<\Omega_m<1$ at the 1\,$\sigma$ confidence
level. If reduced $\chi^2$ method is used, the best-fit
results are $a=48.96\pm0.18$, $b=1.52\pm0.08,$ and
$\Omega_m=0.50\pm0.12$. The second method uses type Ia
supernovae (SNe Ia) to calibrate the $E_{\rm iso}-E_{\rm p}$
correlation. We calibrate 90 high-redshift GRBs in the redshift
range from 1.44 to 8.1. The cosmological constraints from
these 90 GRBs are $\Omega_m=0.23^{+0.06}_{-0.04}$ for flat
$\Lambda$CDM and $\Omega_m=0.18\pm0.11$ and
$\Omega_{\Lambda}=0.46\pm0.51$ for non-flat $\Lambda$CDM.
For the combination of GRB and SNe Ia sample, we obtain
$\Omega_m=0.271\pm0.019$ and $h=0.701\pm0.002$ for the flat
$\Lambda$CDM and  the non-flat $\Lambda$CDM, and the results are
$\Omega_m=0.225\pm0.044$, $\Omega_{\Lambda}=0.640\pm0.082,$ and
$h=0.698\pm0.004$.
These results from calibrated GRBs are consistent
with that of SNe Ia. Meanwhile, the combined data can improve
cosmological constraints significantly, compared to SNe Ia alone.
Our results show that the $E_{\rm iso}-E_{\rm p}$ correlation is
promising to probe the high-redshift universe.}

\keywords{gamma-rays: bursts - cosmology: dark matter - cosmology: dark energy,
type Ia supernovae}

\maketitle

\section{Introduction}

Gamma-ray bursts (GRBs) are the most violent explosions in the
Universe, with the highest isotropic energy up to $10^{54}$ ergs
(for reviews, see M\'{e}sz\'{a}ros 2006; Zhang 2007; Gehrels et al.
2009). Thus, they can be detected to the edge of the visible
Universe \citep{Ciardi2000,Lamb2000,Wang2012}. {For instance,
the spectroscopically confirmed redshift of GRB090423 is about 8.2
\citep{Tanvir2009,Salvaterra2009}. Therefore, they are promising probes for
the high-redshift Universe~\citep[for a recent review,
see][]{Wang14}. Many studies have been carried out to use GRBs for
cosmological purposes, such as the star formation rate
\citep{Totani1997,Wijers1998,Porciani2001,Wang2009,Wang2011a}, the
intergalactic medium metal enrichment \citep{Barkana2004,Wang2012},
dark energy
\citep{Dai2004,Friedman2005,Schaefer2007,Basilakos08,Wang2011b},
reionization \citep{Totani2006,Gallerani2008,Wang2013}, possible
anisotropic acceleration \citep{Wang2014a}, and the two-point correlation
\citep{Li2015}.

To constrain the cosmological parameters, standard rulers or candles
such as baryon acoustic oscillations
\citep[BAO;][]{Cole2005,Eisenstein2005,Anderson2014}, cosmic
microwave background \citep[CMB;][]{Komatsu11,Planck13,Planck15} and
SNe Ia \citep{Riess1998,Perlmutter1999,Suzuki2012} are required. The redshifts of BAO and SNe Ia are low, however, and the CMB is only a
snapshot of cosmic expansion. Some parameters, such as the density
and EOS parameter of dark energy \citep{Wang12,WangD2014,Wang2014b},
might evolve with redshift. GRBs can probe the evolution of these
parameters at high redshifts and serve as complementary tools for
SNe Ia. The study of these evolutions can differentiate dark energy
models. Some luminosity correlations have been proposed to
standardize GRBs \citep{Amati2002,Ghirlanda2004a,Liang2005}.
\cite{Ghirlanda2004a} found a tight correlation between collimated
energy $E_\gamma$ and the peak energy $E_{\rm p}$ of $\nu F_\nu$
spectrum. \cite{Dai2004} used this correlation to constrain
cosmological parameters with 12 GRBs. \cite{Liang2005} found the
$E_{\rm iso}-E_{\rm p}-t_{\rm b}$ correlation and used this
correlation to constrain cosmological parameters. Recently,
\cite{Wang2011b} constrained cosmological parameters with 109 GRBs
using six GRB empirical correlations, and found
$\Omega_m=0.31^{+0.13}_{-0.10}$ in the flat $\Lambda$CDM model.
Other attempts have also been made to standardize GRBs
\citep{Ghirlanda2004b,Friedman2005,Schaefer2007,Wang2007,Liang2008,
Kodama2008,Qi2009,Cardone2010,Wang11}. These methods of standardizing
the long GRBs are mainly based on some empirical correlations, such
as the $E_{\rm iso}-E_{\rm p}$ \citep{Amati2002}, $E_{\rm p}-L_{\rm
p}$ \citep{Schaefer2003,Wei2003}, and $E_{\rm p}-E_{\gamma}$
\citep{Ghirlanda2004a}, where $L_{\rm p}$ is the peak luminosity,
$E_{\rm p}$ is the peak energy in cosmological rest frame, $E_{\rm
iso}$ is the isotropic-equivalent energy, and $E_{\gamma}$ is the
collimation-corrected energy. Correlations within X-ray afterglow
light curves  have also been studied
\citep{Dainotti2008,Dainotti2010,Qi2010}.

In this paper, we focus on the usage of the $E_{\rm iso}-E_{\rm p}$
correlation. \cite{Amati2002} discovered this correlation with a
small sample of $Beppo$SAX GRBs. Since many more GRBs are detected,
attempts have been made to use this correlation for the purpose of cosmology. \cite{Amati2008} used a simultaneous fitting method to
constrain the $E_{\rm iso}-E_{\rm p}$ correlation coefficients and
cosmological parameters together with 70 long GRBs. The extrinsic
scatter $\sigma_{\rm ext}$ was taken into consideration in this
method \citep{D'Agostini2005}. \cite{Amati2008} assigned
$\sigma_{\rm ext}$ to $E_{\rm p}$ and found $0.04<\Omega_m<0.40$ and
$\sigma_{\rm ext}=0.17\pm0.02$ at 1\,$\sigma$ confidence level in the
flat $\Lambda$CDM universe. For non-flat $\Lambda$CDM model, the
results are $\Omega_m\in[0.04, 0.40]$ and $\Omega_{\Lambda}<1.05$
\citep{Amati2008}. However, \cite{Ghirlanda2009} doubted this
result. He claimed that the extrinsic scatter term should be
assigned to $E_{\rm iso}$. This is consistent with
\cite{D'Agostini2005}, who described that the extrinsic scatter
$\sigma_{\rm ext}$ should be assigned to the parameter that also
depends on hidden variables (cosmological parameters in our
study). We  discuss this point in detail in Sect. 3.1.
However, this would lead to no constraint on cosmological parameters
with the same 70 GRBs from \cite{Amati2008}. We test it again with a
larger sample in this paper.

The calibration method is also helpful to standardize GRBs.
Imitating the example of standardizing the standard candle of SNe Ia
with Cepheid variables, GRBs can also be calibrated with SNe Ia
\citep{Liang2008,Kodama2008,Wei2010,Lin2015}. This method is also
cosmological model independent. \cite{Liang2008} calibrated 42 high
redshift GRBs with SNe Ia. Five interpolation methods were used and
the results were consistent with each other. \cite{Wei2010}
standardized 59 high-redshift GRBs with SNe Ia, using the $E_{\rm
iso}-E_{\rm p}$ correlation, and found that GRBs can improve the
constraint on cosmological parameters. \cite{Wang11} calibrated 116
GRBs with Union 2 SNe Ia with cosmographic parameters.

We use 151 GRBs, 109 of which are taken from
\cite{Amati2008} and \cite{Amati2009}. The remaining 42 GRBs are the
updated long GRBs, which were detected by Fermi GBM, Konus-Wind,
Swift-BAT, and Suzaku-WAM. The energy band, fluence, low ($\alpha$),
high ($\beta$) energy photon indices, spectral peak energy, and
redshift are taken from the refined analysis of the corresponding GRB
team. We test whether this larger GRB sample can help to constrain
cosmological models better. First, we constrain the cosmological
parameters and  coefficients of the $E_{\rm iso}-E_{\rm p}$
correlation simultaneously. Then, we calibrate these GRBs with SNe
Ia using the $E_{\rm iso}-E_{\rm p}$ correlation. At last, we
compare these two methods and discuss them.

This paper is organized as follows. In the next section, we introduce
the GRBs data and perform the K-correction. In Sect. 3, we test
whether the redshift evolution of the $E_{\rm iso}-E_{\rm p}$
correlation is significant, and use a simultaneous fitting method to
constrain cosmological parameters and coefficients of the $E_{\rm
iso}-E_{\rm p}$ correlation. In Sect. 4, we use SNe Ia to
calibrate the $E_{\rm iso}-E_{\rm p}$ correlation, then we use these
calibrated GRBs to constrain cosmological parameters. Summary
and discussions are given in Sect. 5.

\section{Updated GRB sample}

We collect all GRBs with information of redshift, fluence,
peak energy, and photon indices from GCN Circulars
Archive\footnote{\url{http://gcn.gsfc.nasa.gov/gcn3_archive.html},}
\cite{Cucchiara2011} and \cite{Gendre2013} until February 13, 2014.
The updated sample contains 42 updated long GRBs.
We list these GRBs in Table \ref{sample}. The
spectra of these GRBs are obtained from the refined analysis of
Fermi GBM team, Konus-Wind team, Swift-BAT team, and Suzaku-WAM team.
The redshifts extend from 0.34 to 5.91. The spectrum is modeled by a
broken power law \citep{Band1993},
\begin{equation}
\Phi(E) = \left \{
\begin{array}{ll}
A E^{\alpha} {\rm e}^{-(2 + \alpha) E/E_{\rm p,obs}} & E \le
\frac{\alpha
-\beta}{2 + \alpha}E_{\rm p,obs} \\ ~ & ~ \\
B E^{\beta} & {\rm otherwise,}
\end{array}
\right.\  \label{band}
\end{equation}
where $E_{\rm p,obs}$ is the observed peak energy, $\alpha$  and
$\beta$ are the low and high energy photon indices, respectively. We
take the typical spectral index values for those GRB whose indices
are not given out in the references, i.e., $\alpha=-1.0$ and
$\beta=-2.2$ \citep{Salvaterra2009}.

With these spectra parameters, we can obtain the peak energy in the
cosmological rest frame by $E_{\rm p}=E_{\rm p,obs}\times (1+z)$ and
the bolometric fluence in the band of $1-10^4$ keV by
\citep{Bloom2001}
\begin{equation}
S_{\rm bolo} = S \ {\times} \ \frac{\int_{1/(1 + z)}^{10^4/(1 +
z)}{E \Phi(E) dE}} {\int_{E_{\rm min}}^{E_{\rm max}}{E \Phi(E) dE}}
\ , \label{sbolo}
\end{equation}
where $S$ is the observed fluence, $E_{\rm min}$ and $E_{\rm max}$
are the detection limits of the instrument, and $z$ is the redshift.

In the $E_{\rm iso}-E_{\rm p}$ plane, $E_{\rm p}$ is an observed
value, which is not dependent on the cosmological model. However,
$E_{\rm iso}$ depends on the cosmological model from
\begin{equation}
E_{\rm iso}=4 \pi d_{\rm L}^2 S_{\rm bolo}(1+z)^{-1},\label{Eiso}
\end{equation}
where $d_{\rm L}$ is the luminosity distance. Assuming a flat
$\Lambda$CDM model, the $d_{\rm L}$ can be expressed with Hubble
expansion rate
\begin{equation}
  d_{\rm L}(\Omega_m,z) = (1+z)\frac{c}{H_0} \int_0^z \frac{dz'}{\sqrt{\Omega_m (1+z)^3 + 1 - \Omega_m}},
  \label{dlflat}
\end{equation}
where $\Omega_m$ is the matter density at present, and $H_0$ is the
Hubble constant. Since the Hubble constant is precisely measured, we
 take $H_0=67.8$ km s$^{-1}$ Mpc$^{-1}$
\citep{Planck13,Planck15}, except when we use the combination data
of SNe and GRB to constrain cosmological models.

We list 42 updated GRBs in Table \ref{sample}. The isotropic energy
$E_{\rm iso}$ is calculated with benchmark parameters with
$\Omega_m=0.308$ for the flat $\Lambda$CDM universe
\citep{Planck13,Planck15}. During the calculation, we only take the errors
propagating from the spectrum parameters, namely observed fluence
$S$ and peak energy $E_{\rm p,obs}$. The uncertainties from other
parameters are attributed into the extrinsic scatter $\sigma_{\rm
ext}$.

\section{The $E_{\rm iso}-E_{\rm p}$ correlation and constraints on cosmological parameters}
\subsection{The $E_{\rm iso}-E_{\rm p}$ correlation}
To constrain cosmological models more precisely, we combine our
updated 42 GRBs with 109 GRBs from \cite{Amati2008} and
\cite{Amati2009}. The full sample contains 151 GRBs and covers the
redshift range from 0.0331 to 8.2. We parameterize the $E_{\rm
iso}-E_{\rm p}$ correlation as follows:
\begin{equation}
\log \frac{E_{\rm iso}}{\text{erg}}=a+b~\log \frac{E_{\rm p}}{\text{keV}},\label{eq:amati}
\end{equation}
where $a$ and $b$ are the intercept and slope. Here $E_{\rm p}$ has been corrected into the cosmological rest frame.

Before constraining cosmological models, we test the possible
redshift evolution of the $E_{\rm iso}-E_{\rm p}$ correlation using
the maximum likelihood method. The full data is divided into four
redshift bins: $[0.0331,0.958]$, $[0.966,1.613]$, $[1.619,2.671],$
and $[2.69,8.2]$. Each bin almost includes the same number of GRBs.
The results are shown in Table \ref{bin}. We give out the best-fit
values and 1\,$\sigma$ uncertainties in the coefficients $a$, $b,$ and
the extrinsic scatter $\sigma_{\rm ext}$. The $\sigma_{\rm ext}$
is almost constant in different bins. Its value is about 0.34, which
implies that the extrinsic scatter dominates the error size. The
results show no statistically significant evidence for the redshift
evolution of the $E_{\rm iso}-E_{\rm p}$ correlation. This result is
consistent with those of \cite{Basilakos08} and \cite{Wang2011b}.
The full data result are also shown in Figure\,\ref{fig:amati}. This  result
illustrates that the $E_{\rm iso}-E_{\rm p}$ correlation fits the
data well.

As discussed by \cite{D'Agostini2005}, we use the following likelihood to fit
the linear relation $y=a+bx$,
\begin{eqnarray}
\label{likelihood} \mathcal{L}(\Omega_m,a,b,\sigma _{\rm ext} ) \propto
\prod\limits_i {\frac{1}{{\sqrt {\sigma ^2 _{ext}   +
\sigma ^2 _{y_i }  + b^2 \sigma ^2 _{x_i } } }}}\;\nonumber\\
\times \exp \left[ - \frac{{(y_i - a - bx_i )^2 }}{{2(\sigma ^2 _{ext}
+ \sigma ^2 _{y_i }  + b^2 \sigma ^2 _{x_i } )}}\right].
\end{eqnarray}
Following the description of \cite{D'Agostini2005}, the parameter
$y$ should not only depend on $x$, but also depend on some hidden
variables ($\Omega_m$ here). Thus, the expression of the $E_{\rm iso}-E_{\rm p}$
plane should be written as $y=\log\frac{E_{\rm
iso}}{\text{erg}}$ and $x=\log\frac{E_{\rm p}}{\text{keV}}$.
However, \cite{Amati2008} set $y=\log\frac{E_{\rm p}}{\text{keV}}$,
thus the extrinsic scatter $\sigma_{\rm ext}$ does not contain the
error from the cosmological models.

\begin{figure}
\begin{center}
\includegraphics[width=0.5\textwidth]{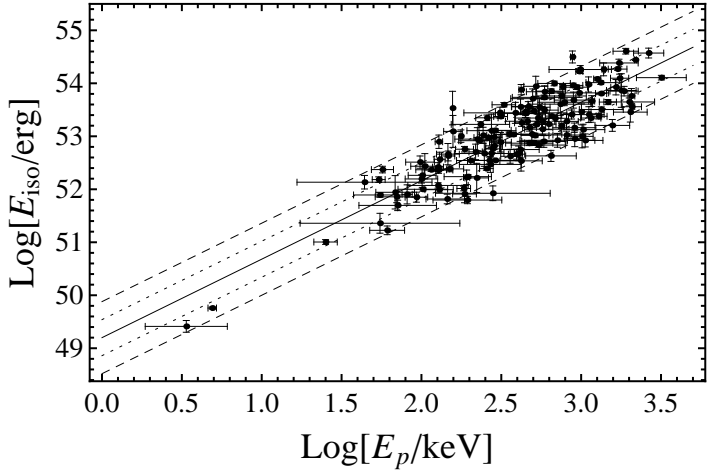}
\end{center}
\caption{The $E_{\rm iso}-E_{\rm p}$ correlation. The solid black,
dotted, and dashed lines represent the best-fit line, $1\,\sigma_{\rm ext}$ region,
and $2\,\sigma_{\rm ext}$ region, respectively. \label{fig:amati}}
\end{figure}

\subsection{Simultaneous fitting}
Since the $E_{\rm iso}-E_{\rm p}$ correlation does not evolve with
redshift, it can be used to constrain parameters directly. We
emphasize that there is no circularity problem in the simultaneous
fitting method because we do not assume any cosmological model. In
this section, we focus on the constraint on the flat $\Lambda$CDM
model. The luminosity distance is expressed as Eq.
(\ref{dlflat}).

Using the likelihood expressed in equation (\ref{likelihood}), we
can constrain the current matter density $\Omega_m$, the extrinsic
scatter parameter $\sigma_{\rm ext}$, and the coefficients of the
$E_{\rm iso}-E_{\rm p}$ correlation simultaneously. In our
calculations, the best-fit values are $a=49.15\pm0.26$,
$b=1.42\pm0.12$, $\sigma_{\rm ext}=0.34\pm0.03,$ and $\Omega_m=0.76$.
We show the constraint on $\Omega_m$ in Fig.\,\ref{fig:om} with
a solid line. The 1\,$\sigma$ uncertainty is
$\Omega_m\in[0.55,1]$. We also use the reduced $\chi^2$
method to constrain the matter density. This method also includes
the effect of extrinsic scatter
\begin{equation}
\chi^2=\sum_i{(y_i - a - bx_i )^2 /(\sigma ^2 _{y_i } + b^2 \sigma
^2 _{x_i } + \sigma^2_{\rm ext})}.
\end{equation}
The hidden variables (cosmological parameters) are included in
$E_{\rm iso}$. The extrinsic scatter is used to set the reduced
$\chi^2$ to unity, which is also used in SNe Ia
cosmology\citep{Suzuki2012}. The value of $\sigma_{\rm ext}$ is 0.34
when the reduced $\chi^2$ is unity. The best-fit results are
$a=48.96\pm0.18$, $b=1.52\pm0.08,$ and $\Omega_m=0.50\pm0.12$. The
constraint from reduced $\chi^2$ method is roughly consistent with
the likelihood method. The $\chi^2/\chi^2_{\rm min}$ evolution with
$\Omega_m$ are shown in Fig. \,\ref{fig:om} with a dashed line. If
the extrinsic scatter is not considered, the results are
$a=48.50\pm0.05$, $b=1.81\pm0.02,$ and $\Omega_m=0.19\pm0.05$.

There is a mild tension between the results from the
likelihood method and the reduced $\chi^2$ method. The extrinsic scatter
is large, which  loosely constrains the cosmological parameters.
When we calculate the parameter
$E_{\rm iso}$, a cosmological model and a spectrum model are used,
while the uncertainties from them are not well established, thus we
take these uncertainties into a scatter parameter $\sigma_{\rm
ext}$. This scatter should be assigned to the parameter $E_{\rm
iso}$. In the future, this scatter can be reduced, since precise
observation and data analysis will be performed by the team of
\emph{Sino-French space-based multiband astronomical variable
objects monitor} \citep[SVOM;][]{Basa2008,Gotz2009,Paul2011}.

We also compare our results to the current precise measurements,
such as the results from $Planck$+WMAP \citep{Planck13}, BAO
\citep{Beutler2011,Anderson2014,Kazin2014,Ross2015}, and SNe Ia
\citep{Conley2011,Suzuki2012}. We show them in Table\,\ref{Tab:om}.
The best-fit $\Omega_m$ by GRBs, using $\chi^2$ method, conflicts with the observation of CMB and BAO.
For the results from SNe Ia, however, if both statistical and
systematic errors are included, the constraints on cosmological
parameters are loose\citep{Kowalski2008,Amanullah2010,Suzuki2012}.
In this case, the best-fit $\Omega_m$ with GRBs, using $\chi^2$ method,
is consistent with those from SNe Ia at $1\sigma$ confidence level;
see Fig. 12 of \cite{Kowalski2008}, Fig. 10 of
\cite{Amanullah2010} and Fig. 5 of \cite{Suzuki2012}.

\begin{figure}
\begin{center}
\includegraphics[width=0.5\textwidth]{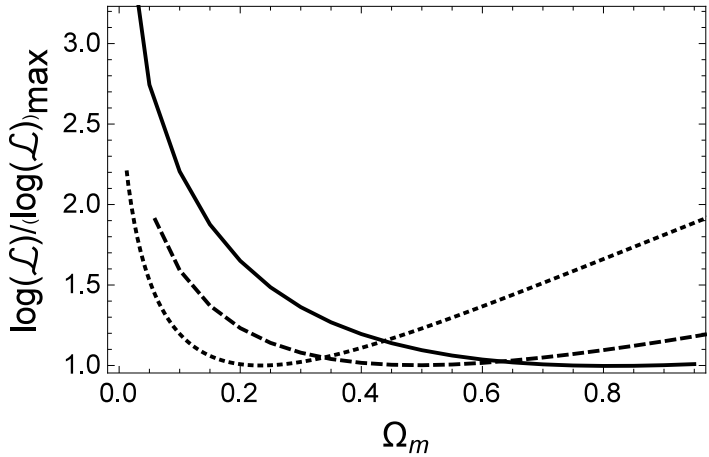}
\end{center}
\caption{The evolution of
$\log(\mathcal{L})/\log(\mathcal{L})_{\rm min}$ as a function of
$\Omega_m$ in the flat $\Lambda$CDM universe is shown with solid
line from maximum likelihood method. The dashed line is the
$\chi^2/\chi^2_{\rm min}-\Omega_m$ plot from reduced $\chi^2$
method. The dotted line is obtained with the 90 GRBs calibrated on the
SNe Ia (see section 4).
\label{fig:om}}
\end{figure}

\section{Calibration of the $E_{\rm iso}-E_{\rm p}$ correlation}
\subsection{Standardizing GRBs with SNe Ia}
Just as using Cepheid variables to standardize SNe Ia, the GRBs
can be calibrated with SNe Ia. We can use the calibrating method
to standardize the GRBs with the $E_{\rm iso}-E_{\rm p}$ correlation.
With this approach, the parameters $a$ and $b$ are obtained and only
cosmological parameters remain free. We use the latest Union 2.1 data
from \cite{Suzuki2012}. This method is also cosmological
model independent \citep{Liang2008,Kodama2008,Wei2010}. The
extrinsic scatter is also be taken into account when calculating
the error propagation of $E_{\rm iso}$. The full GRB
data is separated into two groups. The dividing line is the highest
redshift in SNe Ia Union 2.1 data, namely, $z=1.414$. The
low-redshift group ($z<1.414$) includes 61 GRBs and the
high-redshift group ($z>1.414$) contains 90 GRBs.

Firstly, the linear interpolation method is used to calibrate the
distance moduli $\mu$ of 61 low-redshift GRBs. \cite{Liang2008} have
shown that there are no differences on the final result between the
linear interpolation and the cubic interpolation. The 1\,$\sigma$
error of the distance moduli $\sigma_{\mu,i}$ can be obtained as
follows:\begin{equation}
\sigma_{\mu}^2=(\frac{z_{i+1}-z}{z_{i+1}-z_i})^2\epsilon_{\mu,i}^2+
(\frac{z-z_{i}}{z_{i+1}-z_i})^2\epsilon_{\mu,i+1}^2,
\end{equation}
where $z_{i+1}$ and $z_i$ are the redshift of the two nearest SNe Ia and
$\epsilon_{\mu,i+1}$ and $\epsilon_{\mu,i}$ are the errors of these two
SNe Ia. The redshift of interpolated GRB lies between $z_{i}$ and $z_{i+1}$.

After the distance moduli of 61 low-redshift GRBs are obtained, the
luminosity distance can be derived from
\begin{equation}
\mu=5\log \frac{d_{\rm L}}{\text{Mpc}}+25.\label{dltomu}
\end{equation}
Then the isotropic-equivalent energy $E_{\rm iso}$ can be calculated
from Eq. (\ref{Eiso}). Following \cite{Schaefer2007} and
\cite{Liang2008}, we use the bisector of the two ordinary least squares method \citep{Isobe1990} to fit the $E_{\rm iso}-E_{\rm p}$
correlation. The best-fit values are $a=48.46\pm0.033$ and
$b=1.766\pm0.007$. The result is shown in Fig.
\ref{fig:lowzAmati}. The errors of distance moduli are not taken
into consideration because the extrinsic scatter $\sigma_{\rm
ext}$ dominates the error size in the regression analysis
\citep{Schaefer2007}. Thus, we take $\sigma_{\rm ext}$ directly into account
during the calculations of the uncertainties of high-redshift GRBs
($\sigma_{\log E_{\rm iso}}$). From the previous section,
the value of $\sigma_{\rm ext}$ is nearly constant, so we typically set
$\sigma_{\rm ext}=0.34$.

\begin{figure}
\begin{center}
\includegraphics[width=0.5\textwidth]{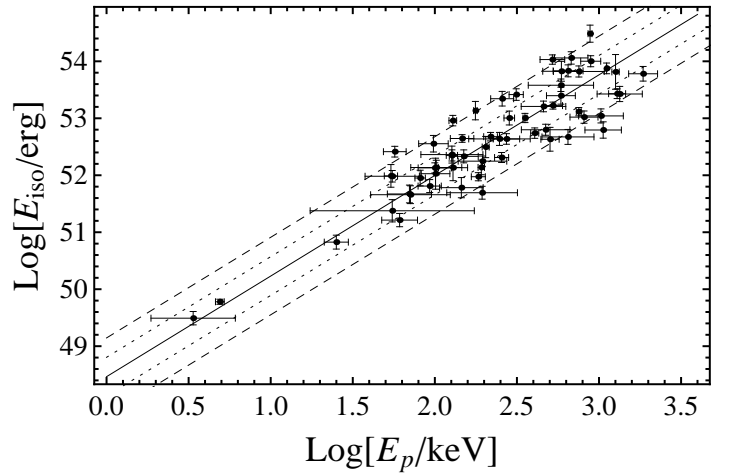}
\end{center}
\caption{Low-redshift GRM sample $E_{\rm iso}-E_{\rm p}$ correlation. Black line is the best-fit result obtained by using the
bisector of the two ordinary least squares method. The dotted line represents the $1\,\sigma_{\rm ext}$ region
and dashed line the
$2\,\sigma_{\rm ext}$ region.
\label{fig:lowzAmati}}
\end{figure}

We have shown that the $E_{\rm iso}-E_{\rm p}$ correlation
does not evolve with redshift in the previous section. Thus, the
calibrated $E_{\rm iso}-E_{\rm p}$ correlation can be extrapolated
to the high-redshift sample, namely, $z>1.414$ group. Using Eq.
(\ref{eq:amati}), we can derive $E_{\rm iso}$ of high-redshift GRBs.
The propagated uncertainties of $E_{\rm iso}$ can be calculated from
\begin{equation}
\sigma_{\log E_{\rm iso}}^2=\sigma_a^2+\left(\sigma_b
\log\frac{E_{\rm p}}{\text{keV}}\right)^2+\left(
\frac{b}{\ln 10}\,\frac{\sigma_{E_{\rm p}}}{E_{\rm p}}\right)^2
+\sigma_{\rm ext}^2,
\end{equation}
where the value of $\sigma_{\rm ext}$ is 0.34. The values of
$\sigma_a$ and $\sigma_b$ are derived from the bisector of the two
ordinary least squares method.

Then, we use Eq. (\ref{Eiso}) and Eq. (\ref{dltomu}) to
derive the distance moduli. The propagated uncertainty is given by
the following equation:
\begin{equation}
\sigma_\mu=\left[\left(\frac{5}{2}\sigma_{\log E_{\rm iso}}\right)^2
+\left(\frac{5}{2\ln 10}\,\frac{\sigma_{S_{\rm bolo}}}{S_{\rm bolo}}
\right)^2\right]^{1/2}.
\end{equation}

The calibrated 90 high-redshift GRBs  are listed in Table
\ref{calibratedGRB}. This sample can be used to constrain
cosmological models directly. Compared with \cite{Wei2010}, the
error bars of distance moduli of our results are smaller. The main
reason is that we use a larger sample, which leads to a smaller
$\sigma_{\rm ext}$. The extrinsic scatter parameter has been
taken into consideration during the calculation of the error size of
$E_{\rm iso}$.

\subsection{Constraining cosmological models}

These GRBs carry the information of high-redshift universe, and can
be taken as good complements to the Union 2.1 data set. We
test if these high-redshift GRBs alone can constrain the
$\Lambda$CDM model. Using the distance modulus in Eq.
(\ref{dlflat}) and Eq. (\ref{dltomu}), the $\chi^2$ is
\begin{equation}
\chi^2_{\text{GRB}}({\Omega_m})=\sum\limits_{i=1}^{90}\frac{\left[
\mu_{\rm cal}(z_i)-\mu(z_i)\right]^2}{\sigma^2(z_i)}\,,
\end{equation}
where $\mu_{\rm cal}$ is the calibrated GRB distance
modulus listed in Table \ref{calibratedGRB}. The best-fit result is
$\Omega_m=0.23^{+0.06}_{-0.04}$ with 1\,$\sigma$ uncertainty. The
$\chi^2$ evolution with $\Omega_m$ is shown in Fig. \,\ref{fig:om}.
This result is consistent with the constraints from
SNe Ia \citep{Conley2011,Suzuki2012}, CMB \citep{Planck13,Planck15},
and BAO \citep{Beutler2011,Anderson2014,Kazin2014,Ross2015} at
1\,$\sigma$ confidence level, as shown in Table \,\ref{Tab:om}.

Since this GRB sample can constrain cosmological parameters successfully,
we also combine the calibrated GRB data with SNe Ia from Union 2.1 sample
to constrain cosmological models. For the flat $\Lambda$CDM, we obtain
$\Omega_m=0.271\pm0.019$ and $h=0.701\pm0.002$, where $h$ is the Hubble constant
in units of 100 km s$^{-1}$ Mpc$^{-1}$. This is very consistent with  the Union 2.1 SNe Ia
data. For the non-flat $\Lambda$CDM,
the luminosity distance is different and can be expressed as follows:
\begin{equation}
d_{L}=\left\{
\begin{array}{l}
\displaystyle
cH_{0}^{-1}(1+z)(-\Omega_{k})^{-1/2}\sin[(-\Omega_{k})^{1/2}I],
~ \Omega_{k}<0, \\
\displaystyle cH_{0}^{-1}(1+z)I,
~~~~~~~~~~~~~~~~~~~~~~~~~~~~~~~~~~~~~~~~~
\Omega_{k}=0,\\
\displaystyle
cH_{0}^{-1}(1+z)\Omega_{k}^{-1/2}\sinh[\Omega_{k}^{1/2}I],
~~~~~~~~~~~~~~\,  \Omega_{k}>0,\\
\end{array} \right.
\label{eqn:fc:}
\end{equation}
where
\begin{equation}
\Omega_{k}=1-\Omega_{m}-\Omega_\Lambda,
\end{equation}
and \begin{equation}
I=\int_{0}^{z}\frac{dz}{\sqrt{(1+z)^{3}\Omega_{m}+\Omega_\Lambda+(1+z)^{2}\Omega_{k}}}.
\end{equation}
The $\chi^2$ of SNe Ia is constructed as follows:
\begin{equation}
\chi^2_{\text{SNe}}({h, \Omega_m,\Omega_{\Lambda}})=\sum\limits_{i=1}^{580}\frac{\left[
\mu_{\rm obs}(z_i)-\mu(z_i)\right]^2}{\sigma^2(z_i)}.
\end{equation}
Then the total $\chi^2$ is
\begin{equation}
\chi^2_{\text{total}}({h, \Omega_m,\Omega_{\Lambda}})
=\chi^2_{\text{SNe}}({h, \Omega_m,\Omega_{\Lambda}})
+\chi^2_{\text{GRB}}({h, \Omega_m,\Omega_{\Lambda}}).
\end{equation}

The best-fit values with 1\,$\sigma$ uncertainties are
$\Omega_m=0.225\pm0.044$, $\Omega_{\Lambda}=0.640\pm0.082,$ and
$h=0.698\pm0.004$ for the combined sample (SNe+GRB). For the GRB
sample, we obtain $\Omega_m=0.18\pm0.11$ and
$\Omega_{\Lambda}=0.46\pm0.51$, which is consistent with the SNe Ia
results at 1\,$\sigma$ confidence level. The combined sample can help
to constrain cosmological parameters much tighter because not only is
the sample enlarged, but also the redshift covers a much wider. The
flatness of the Universe depends on the curvature parameter, that is to say,
$\Omega_{k}=1-\Omega_{\Lambda}-\Omega_m$. In Fig.\,\ref{fig:omde},
we use three samples, GRB, SNe, and combination of GRB+SNe
to constrain the cosmological model. Both results prefer a flat
universe at the 1\,$\sigma$ confidence level. The constraint
from the GRB is almost perpendicular to that from SNe Ia in the
$\Omega_m-\Omega_{\Lambda}$ plane. Thus GRBs can significantly help
to constrain  $\Omega_m$ because,  in this
redshift domain, the dark matter dominates the evolution of the
Universe. We also show constraints on $\Omega_m-h$ in Fig.
\ref{fig:omh}, and $\Omega_{\Lambda}-h$ in Fig. \,\ref{fig:deh}.

\begin{figure}
\begin{center}
\includegraphics[width=0.5\textwidth]{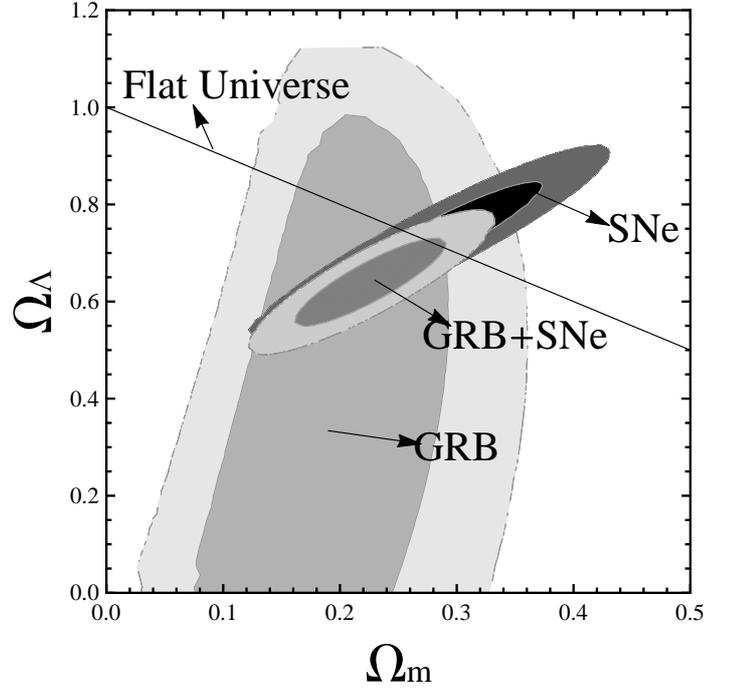}
\end{center}
\caption{1\,$\sigma$ and 2$\sigma$ constraints on $\Omega_m$ and
$\Omega_{\Lambda}$. We use three samples and plot them into different colors.
The solid line shows the $\Omega_k=0$ case.
\label{fig:omde}}
\end{figure}

\begin{figure}
\begin{center}
\includegraphics[width=0.5\textwidth]{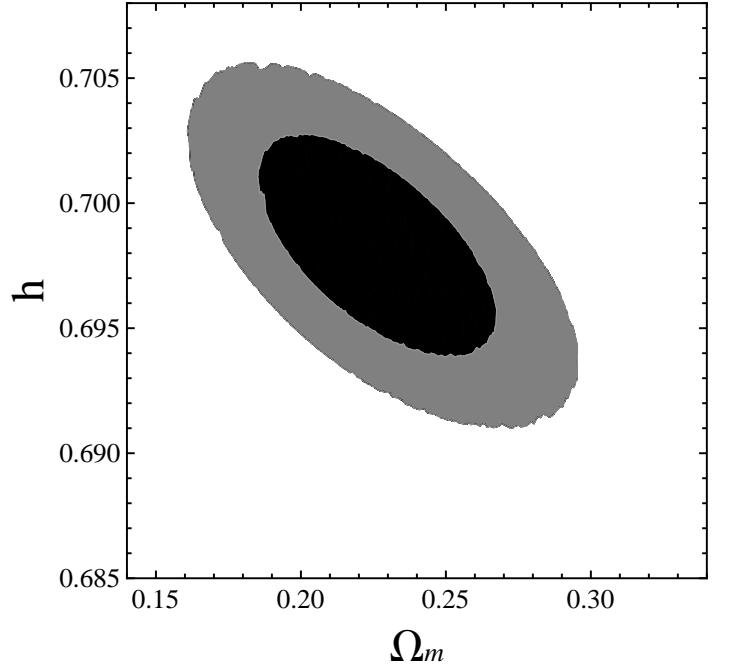}
\end{center}
\caption{1\,$\sigma$ and 2$\sigma$ constraints on $\Omega_m$ and
$h$ from SNe Ia and GRB data. \label{fig:omh}}
\end{figure}

\begin{figure}
\begin{center}
\includegraphics[width=0.5\textwidth]{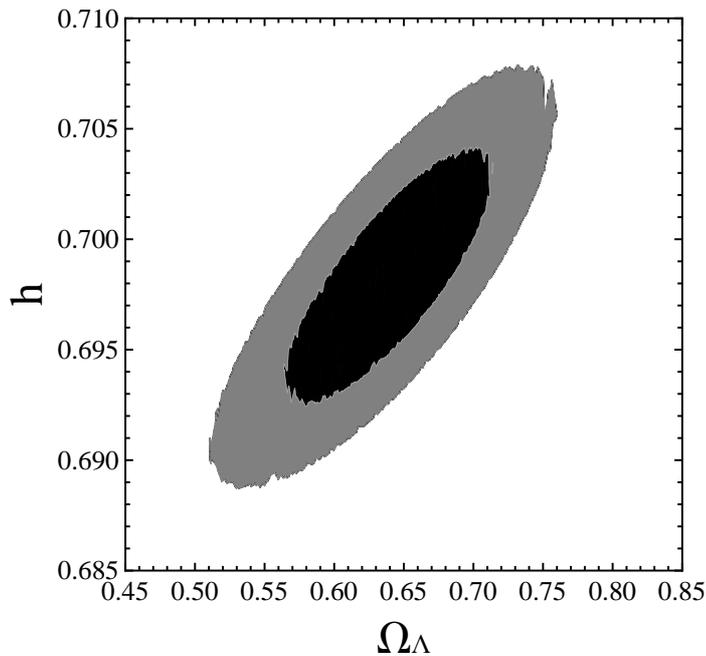}
\end{center}
\caption{1\,$\sigma$ and 2$\sigma$ constraints on the
$\Omega_{\Lambda}$ and $h$ from SNe Ia and GRB data.
\label{fig:deh}}
\end{figure}

\section{Discussions and summary}

In this paper, we update 42 long GRBs for the $E_{\rm
iso}-E_{\rm p}$ correlation and combine them with 109 long GRBs from
\cite{Amati2008} and \cite{Amati2009}. This sample contains GRBs
detected by different detectors with different sensitivities.
Thus, the sample might be biased, but this bias should only have a weak effect on
our results. We also use the complete sample to perform our
analysis. We use the same criteria as Salvaterra et al. (2012) and
Pescalli et al. (2015) to collect GRBs. The results are
$a=49.45\pm0.61$, $b=1.24\pm0.22$ and $\sigma_{\rm ext}=0.38\pm0.06$,
while no constraint on $\Omega_m$ is found. These results are in
tension with that of our updated full sample with a larger extrinsic
scatter. No statistical evidence for the redshift evolution of the
$E_{\rm iso}-E_{\rm p}$ is found in the full sample.

For cosmological purposes, we fit the $E_{\rm iso}-E_{\rm p}$ plane
and the cosmological parameters simultaneously. Using a likelihood
function we obtain $a=49.15\pm0.26$, $b=1.42\pm0.11$, $\sigma_{\rm
ext}=0.34\pm0.03,$ and $\Omega_m\in[0.55,1]$. Using the
reduced $\chi^2$, we obtain $a=48.96\pm0.18$, $b=1.52\pm0.08,$ and
$\Omega_m=0.50\pm0.12$. The results from these two fitting methods
are in mild tension. The main reason is that the extrinsic scatter
of this correlation is too large. Thus, \cite{Ghirlanda2009} finds no
constraint with a smaller sample using the likelihood method. We also use a
calibrating method. Based on the SNe Ia data, we
obtain 90 calibrated GRBs. From these calibrated GRBs, we acquire
$\Omega_m=0.23^{+0.06}_{-0.04}$ for flat $\Lambda$CDM and for the
non-flat $\Lambda$CDM, we obtain $\Omega_m=0.18\pm0.11$ and
$\Omega_{\Lambda}=0.46\pm0.51$. We also combine the GRB sample with
SNe Ia Union\,2.1 data and obtain $\Omega_m=0.271\pm0.019$ and
$h=0.701\pm0.002$ for the flat $\Lambda$CDM. For the non-flat
$\Lambda$CDM, the results are $\Omega_m=0.225\pm0.044$,
$\Omega_{\Lambda}=0.640\pm0.082,$ and $h=0.698\pm0.004$. We list our
results in Table\,\ref{Tab:om}, and compare them with the results
from other current measurements. The results from GRBs are
consistent with results from SNe Ia in 1\,$\sigma$ confidence level
\citep{Conley2011,Suzuki2012}, while they conflict with CMB
\citep{Planck13,Planck15} and BAO
\citep{Beutler2011,Anderson2014,Kazin2014,Ross2015}. We also found
that the GRBs can help to constrain dark matter better. The
constraint from GRB are almost perpendicular to that from SNe Ia in
the $\Omega_m-\Omega_{\Lambda}$ plane. The main reason might be that
at high redshift, the dark matter dominates the Universe.

The extrinsic scatter is taken into account in both the
simultaneous fitting method and the calibrating method. Our results
shows that tighter constraints on cosmological model can be obtained
with the calibrating method. For the simultaneous fitting
method, the reduced $\chi^2$ method gives a more stringent constraint
on cosmological parameters than the likelihood method, but the constraint
is still loose because of the large extrinsic scatter. This scatter is
introduced by both cosmological models and the GRB spectrum parameters,
such as $E_{\rm p}$, fluence, and photon index. The spectrum parameters can be
precisely measured by SVOM \citep{Basa2008,Gotz2009,Paul2011}, which
can reduce the extrinsic scatter. The GRBs from SVOM would better help shed
light on the properties of early Universe.

\section*{Acknowledgements}
We thank an anonymous referee for useful suggestions and comments.
This work is supported by the National Basic Research Program of
China (973 Program, grant No. 2014CB845800), the National Natural
Science Foundation of China (grants 11422325, 11373022, 11103007,
and 11033002), the Excellent Youth Foundation of Jiangsu Province
(BK20140016), and the Program for New Century Excellent Talents in
University (grant No. NCET-13-0279). KSC is supported by the CRF
Grants of the Government of the Hong Kong SAR under HUKST4/CRF/13G.

\newpage
\onecolumn
{\small
\begin{longtable}{@{} c @{ } c @{ } c @{ } c @{ } c @{ } c @{ } c }
 \hline
 GRB    &   Redshift &  $S_{\rm bolo}$($10^{-5}$ erg cm$^{-2}$)   &  $E_{\rm p}$(keV)  &  $E_{\rm iso}^{(a)}$($10^{53}$ erg) & Instruments$^{(b)}$    &  Refs. for spectrum$^{(c)}$\\
\hline
\endhead
100413  &   3.90    &    $  2.36    \pm 0.77    $  & $  1783.60     \pm 374.85  $  & $  7.31    \pm 4.56     $  &   SW  &   (1) \\
100621  &   0.54    &    $  5.75    \pm 0.64    $  & $  146.49  \pm 23.90   $  & $  0.46    \pm 0.20    $  &     KW &   (2) \\
100704  &   3.60    &    $  0.70    \pm 0.07    $  & $  809.60  \pm 135.70  $  & $  1.91    \pm 0.61    $  &     KW &   (3) \\
100728B &   2.45    &    $  0.29    \pm 0.01    $  & $  359.11  \pm 48.34   $  & $  0.42    \pm 0.12    $  &     FG &   (4) \\
100814  &   1.44    &    $  1.39    \pm 0.23    $  & $  312.32  \pm 48.80   $  & $  0.77    \pm 0.31    $  &     KW &   (5) \\
100906  &   1.73    &    $  3.56    \pm 0.55    $  & $  387.23  \pm 244.07  $  & $  2.77    \pm 1.18    $  &     KW &   (6) \\
110205  &   2.22    &    $  3.32    \pm 0.68    $  & $  740.60  \pm 322.00  $  & $  4.04    \pm 1.82    $  &     KW/SB/SW   &   (7) \\
110213  &   1.46    &    $  1.55    \pm 0.23    $  & $  223.86  \pm 70.11   $  & $  0.88    \pm 0.41    $  &     KW &   (8) \\
110422  &   1.77    &    $  9.32    \pm 0.02    $  & $  421.04  \pm 13.85   $  & $  7.58    \pm 1.67    $  &     KW &   (9) \\
110503  &   1.61    &    $  2.76    \pm 0.21    $  & $  572.25  \pm 50.95   $  & $  1.89    \pm 0.55    $  &     KW &   (10)    \\
110715  &   0.82    &    $  2.73    \pm 0.24    $  & $  218.40  \pm 20.93   $  & $  0.51    \pm 0.16    $  &     KW &   (11)    \\
110731  &   2.83    &    $  2.51    \pm 0.01    $  & $  1164.32     \pm 49.79   $  & $  4.62    \pm 1.06     $  &   KW  &   (12)    \\
110818  &   3.36    &    $  1.05    \pm 0.08    $  & $  1117.47     \pm 241.11  $  & $  2.56    \pm 0.85     $  &   FG  &   (13)    \\
111008  &   5.00    &    $  1.06    \pm 0.11    $  & $  894.00  \pm 240.00  $  & $  4.82    \pm 1.61    $  &     KW &   (14)    \\
111107  &   2.89    &    $  0.18    \pm 0.03    $  & $  420.44  \pm 124.58  $  & $  0.34    \pm 0.14    $  &     FG &   (15)    \\
111209  &   0.68    &    $  69.47   \pm 8.72    $  & $  519.87  \pm 88.88   $  & $  8.77    \pm 3.61    $  &     KW &   (16)    \\
120119  &   1.73    &    $  4.62    \pm 0.59    $  & $  417.38  \pm 54.56   $  & $  3.60    \pm 1.17    $  &     KW &   (17)    \\
120326  &   1.80    &    $  0.44    \pm 0.02    $  & $  129.97  \pm 10.27   $  & $  0.37    \pm 0.11    $  &     FG &   (18)    \\
120724  &   1.48    &    $  0.15    \pm 0.02    $  & $  68.45   \pm 18.60   $  & $  0.09    \pm 0.05    $  &     SB &   (19)    \\
120802  &   3.80    &    $  0.43    \pm 0.07    $  & $  274.33  \pm 93.04   $  & $  1.28    \pm 0.78    $  &     SB &   (20)    \\
120811C &   2.67    &    $  0.74    \pm 0.07    $  & $  157.49  \pm 20.92   $  & $  1.24    \pm 0.74    $  &     SB &   (21)    \\
120909  &   3.93    &    $  2.69    \pm 0.23    $  & $  1651.55     \pm 123.25  $  & $  8.44    \pm 2.72     $  &   KW  &   (22)    \\
120922  &   3.10    &    $  1.59    \pm 0.18    $  & $  156.62  \pm 0.04    $  & $  3.41    \pm 2.12    $  &     SB &   (23)    \\
121128  &   2.20    &    $  0.87    \pm 0.07    $  & $  243.20  \pm 12.80   $  & $  1.04    \pm 0.35    $  &     KW &   (24)    \\
130215  &   0.60    &    $  4.84    \pm 0.12    $  & $  247.54  \pm 100.61  $  & $  0.47    \pm 0.24    $  &     FG &   (25)    \\
130408  &   3.76    &    $  0.99    \pm 0.17    $  & $  1003.94     \pm 137.98  $  & $  2.89    \pm 0.96     $  &   KW  &   (26)    \\
130420A &   1.30    &    $  1.73    \pm 0.06    $  & $  128.63  \pm 6.89    $  & $  0.79    \pm 0.22    $  &     FG &   (27)    \\
130427A &   0.34    &    $  311.17  \pm 0.47    $  & $  1112.20     \pm 6.70    $  & $  9.51    \pm 3.01     $  &   FG  &   (28)    \\
130505  &   2.27    &    $  4.56    \pm 0.09    $  & $  2063.37     \pm 101.37  $  & $  5.77    \pm 1.79     $  &   KW  &   (29)    \\
130514  &   3.60    &    $  1.88    \pm 0.25    $  & $  496.80  \pm 151.80  $  & $  5.13    \pm 2.05    $  &     KW/SB  &   (30)    \\
130518  &   2.49    &    $  12.34   \pm 0.08    $  & $  1382.04     \pm 31.41   $  & $  18.31   \pm 4.97     $  &   FG  &   (31)    \\
130606  &   5.91    &    $  0.49    \pm 0.09    $  & $  2031.54     \pm 483.70  $  & $  2.86    \pm 1.16     $  &   KW  &   (32)    \\
130610  &   2.09    &    $  0.82    \pm 0.05    $  & $  911.83  \pm 132.65  $  & $  0.90    \pm 0.30    $  &     FG &   (33)    \\
130612  &   2.01    &    $  0.08    \pm 0.01    $  & $  186.07  \pm 31.56   $  & $  0.08    \pm 0.03    $  &     FG &   (34)    \\
130701A &   1.16    &    $  0.46    \pm 0.04    $  & $  191.80  \pm 8.62    $  & $  0.17    \pm 0.05    $  &     KW &   (35)    \\
130831A &   0.48    &    $  1.29    \pm 0.07    $  & $  81.35   \pm 5.92    $  & $  0.08    \pm 0.03    $  &     KW &   (36)    \\
130907A &   1.24    &    $  75.21   \pm 4.76    $  & $  881.77  \pm 24.62   $  & $  31.40   \pm 7.97    $  &     KW &   (37)    \\
131030A &   1.29    &    $  1.05    \pm 0.10    $  & $  405.86  \pm 22.93   $  & $  0.48    \pm 0.15    $  &     KW &   (38)    \\
131105A &   1.69    &    $  4.75    \pm 0.16    $  & $  547.68  \pm 83.53   $  & $  3.54    \pm 1.28    $  &     FG &   (39)    \\
131117A &   4.04    &    $  0.05    \pm 0.01    $  & $  221.85  \pm 37.31   $  & $  0.16    \pm 0.09    $  &     SB &   (40)    \\
140206A &   2.73    &    $  1.69    \pm 0.03    $  & $  447.60  \pm 22.38   $  & $  2.93    \pm 0.74    $  &     FG &   (41)    \\
140213A &   1.21    &    $  2.53    \pm 0.04    $  & $  176.61  \pm 4.42    $  & $  1.01    \pm 0.26    $  &     FG &   (42)    \\
\hline
\caption{\label{sample} 42 updated long GRBs.\\
(a) $E_{\rm iso}$ is computed with benchmark
parameters: $H_0=67.8$ km s$^{-1}$ Mpc$^{-1}$ and $\Omega_m=0.308$; \\
(b) Instruments: FG=Fermi GBM, KW=Konus-Wind, SB=
Swift-BAT and SW=Suzaku-WAM;\\
(c) References for the spectrum parameters: (1) Sugita et al.
(2010); (2) Golenetskii et al. (2010a); (3) Golenetskii et al.
(2010b); (4) von Kienlin et al. (2010); (5) Golenetskii et al.
(2010c); (6) Golenetskii et al. (2010d); (7) Cucchiara et al.
(2011); (8) Golenetskii et al. (2011a); (9) Golenetskii et al.
(2011b); (10) Golenetskii et al. (2011c); (11) Golenetskii et al.
(2011d); (12) Golenetskii et al. (2011e); (13) Xiong et al. (2011);
(14) Golenetskii et al. (2011f); (15) Pelassa et al. (2011); (16)
Golenetskii et al. (2011g); (17) Golenetskii et al. (2012a); (18)
Collazzi et al. (2012); (19) Krimm et al. (2012a); (20) Stamatikos
et al. (2012); (21) Krimm et al. (2012b); (22) Golenetskii et al.
(2012b); (23) Krimm et al. (2012c); (24) Golenetskii et al. (2013a)
(25) Younes et al. (2013); (26) Golenetskii et al. (2013b); (27)
Xiong et al. (2013a); (28) von Kienlin et al. (2013); (29)
Golenetskii et al. (2013c); (30) Palshin et al. (2013); (31) Xiong
et al. (2013b); (32) Golenetskii et al. (2013d) (33) Fitzpatrick et
al. (2013a); (34) Fitzpatrick et al. (2013b); (35) Golenetskii et
al. (2013e); (36) Golenetskii et al. (2013f) (37) Golenetskii et al.
(2013g); (38) Golenetskii et al. (2013h); (39) Fitzpatrick et al.
(2013c); (40) Krimm et al. (2013); (41) von Kienlin et al. (2014);
(42) Zhang (2014).} \centering
\end{longtable}
}

\newpage
\begin{table}
\centering
\begin{tabular}{c|c|c|c|c}
\hline
Redshift range &$a $ & $b$ & $\sigma_{\rm ext}$ &  GRB number \\
\hline
Full data  &   $ 49.21\pm0.24    $   &   $  1.48\pm 0.09    $   &   $    0.34\pm0.04   $&   151\\
$[0.0331,0.958]$ &   $   48.92\pm0.36 $   &   $  1.58\pm0.15    $  &   $   0.34\pm0.07   $   &   37 \\
$[0.966,1.613]$  &   $  49.54\pm0.61$   &   $ 1.37\pm0.23    $  &   $  0.40\pm0.07  $   &  38 \\
$[1.619,2.671]$   &   $  49.62\pm0.64   $   &   $1.33\pm0.24   $   &  $ 0.33\pm0.07 $   &  38 \\
$[2.69,8.1]$ &   $  49.62\pm0.60 $   &   $ 1.34\pm0.22  $   &   $  0.25\pm0.07   $&   38 \\
\hline
\end{tabular}
\caption{The $E_{\rm iso}-E_{\rm p}$ correlation fitting results of full
data and four redshift bins. The best-fit value, 1\,$\sigma$
uncertainties, and extrinsic scatter $\sigma_{\rm ext}$ are given.}
\label{bin}
\end{table}

\begin{table}
\centering
\begin{tabular}{c|c|c|c}
\hline
Data & Cosmological model & Constraint & Method \\
\hline
GRB & flat $\Lambda$CDM & $\Omega_m\in[0.55,1]$ & simultaneous
fitting by likelihood\\
GRB & flat $\Lambda$CDM &$\Omega_m=0.50\pm0.12 $& simultaneous
fitting by $\chi^2$ \\
GRB & flat $\Lambda$CDM &$\Omega_m=0.23_{-0.04}^{+0.06} $&calibrated on the SNe Ia \\
GRB & non-flat $\Lambda$CDM &$\Omega_m=0.18\pm0.11$, $\Omega_{\Lambda}=0.46\pm0.51$ &calibrated on the SNe Ia\\
SNe + GRB & flat $\Lambda$CDM&$\Omega_m=0.271\pm0.019$& calibrated on the SNe Ia\\
SNe + GRB & non-flat $\Lambda$CDM &$ \Omega_m=0.225\pm0.044$, $\Omega_{\Lambda}=0.640\pm0.082$ & calibrated on the SNe Ia\\
$Planck$+BAO & flat $\Lambda$CDM &$\Omega_m=0.315_{-0.018}^{+0.016}$& \\
SNe Union\,2.1& flat $\Lambda$CDM &$\Omega_m=0.277\pm0.022 $& \\
\hline
\end{tabular}
\caption{The constraints of cosmological parameters by GRBs. Simultaneous fitting and calibrating methods are used. We also show the constraints
of cosmological parameters with other measurements for comparison. }
\label{Tab:om}
\end{table}

\newpage
\onecolumn
{\tiny
\begin{longtable}{@{} c @{ } c @{ } c @{ } c @{ } c  |@{ } c @{} c @{ } c @{ } c @{ } c @{ } c }
 \hline
 GRB    &   Redshift &  $S_{\rm bolo} $   &  $E_{\rm p}$  &  $\mu_{\rm cal}$ & GRB    &   Redshift &  $S_{\rm bolo} $ &  $E_{\rm p}$  &  $\mu_{\rm cal}$\\
& z& ($10^{-5}$ erg cm$^{-2}$) & (keV) & & & z& ($10^{-5}$ erg cm$^{-2}$)& (keV)& \\
\hline
050318  &   1.44    &    $  0.42    \pm 0.03    $  & $  115 \pm 25  $  & $  44.48   \pm 1.28    $  &     130518 &   2.49    &    $  12.34   \pm 0.08    $  & $  1382.04     \pm 31.41   $  & $  45.96   \pm 0.86     $      \\
100814  &   1.44    &    $  1.39    \pm 0.23    $  & $  312.32  \pm 48.80   $  & $  45.09   \pm 1.11    $  &     081121 &   2.512   &    $  1.71    \pm 0.33    $  & $  871 \pm 123 $  & $  47.23   \pm 1.08    $   \\
110213  &   1.46    &    $  1.55    \pm 0.23    $  & $  223.86  \pm 70.11   $  & $  44.34   \pm 1.63    $  &     081118 &   2.58    &    $  0.27    \pm 0.057   $  & $  147 \pm 14  $  & $  45.84   \pm 0.98    $   \\
010222  &   1.48    &    $  14.6    \pm 1.5 $  & $  766 \pm 30  $  & $  44.28   \pm 0.88    $  &    080721   &  2.591   &    $  7.86    \pm 1.37    $  & $  1741    \pm 227 $  & $  46.92   \pm 1.04    $   \\
120724  &   1.48    &    $  0.15    \pm 0.02    $  & $  68.45   \pm 18.60   $  & $  44.62   \pm 1.48    $  &     050820 &   2.612   &    $  6.4 \pm 0.5 $  & $  1325    \pm 277 $  & $  46.63   \pm 1.26    $   \\
060418  &   1.489   &    $  2.3 \pm 0.5 $  & $  572 \pm 143 $  & $  45.73   \pm 1.41    $  &    030429  &    2.65   &    $  0.14    \pm 0.02    $  & $  128 \pm 26  $  & $  46.31   \pm 1.25    $   \\
030328  &   1.52    &    $  6.4 \pm 0.6 $  & $  328 \pm 55  $  & $  43.56   \pm 1.13    $  &    120811C &    2.671  &    $  0.74    \pm 0.07    $  & $  157.49  \pm 20.92   $  & $  44.91   \pm 1.04    $   \\
070125  &   1.547   &    $  13.3    \pm 1.3 $  & $  934 \pm 148 $  & $  44.79   \pm 1.11    $  &    080603B  &  2.69    &    $  0.64    \pm 0.058   $  & $  376 \pm 100 $  & $  46.74   \pm 1.45    $   \\
090102  &   1.547   &    $  3.48    \pm 0.63    $  & $  1149    \pm 166 $  & $  46.64   \pm 1.08    $  &     140206A    &   2.73    &    $  1.69    \pm 0.03    $  & $  447.60  \pm 22.38   $  & $  46.03   \pm 0.88    $    \\
040912  &   1.563   &    $  0.21    \pm 0.06    $  & $  44  \pm 33  $  & $  43.44   \pm 3.43    $  &     091029 &   2.752   &    $  0.47    \pm 0.044   $  & $  230 \pm 66  $  & $  46.15   \pm 1.53    $   \\
990123  &   1.6 &    $  35.8    \pm 5.8 $  & $  1724    \pm 466 $  & $  44.91   \pm 1.48    $  &    081222   &  2.77    &    $  1.67    \pm 0.17    $  & $  505 \pm 34  $  & $  46.29   \pm 0.91    $   \\
071003  &   1.604   &    $  5.32    \pm 0.59    $  & $  2077    \pm 286 $  & $  47.34   \pm 1.05    $  &     050603 &   2.821   &    $  3.5 \pm 0.2 $  & $  1333    \pm 107 $  & $  47.36   \pm 0.92    $   \\
090418  &   1.608   &    $  2.35    \pm 0.59    $  & $  1567    \pm 384 $  & $  47.69   \pm 1.40    $  &     110731 &   2.83    &    $  2.51    \pm 0.01    $  & $  1164.32     \pm 49.79   $  & $  47.46   \pm 0.87     $      \\
110503  &   1.613   &    $  2.76    \pm 0.21    $  & $  572.25  \pm 50.95   $  & $  45.58   \pm 0.94    $  &     111107 &   2.893   &    $  0.18    \pm 0.03    $  & $  420.44  \pm 124.58  $  & $  48.39   \pm 1.57    $    \\
990510  &   1.619   &    $  2.6 \pm 0.4 $  & $  423 \pm 42  $  & $  45.07   \pm 0.97    $  &    050401  &    2.9    &    $  1.9 \pm 0.4 $  & $  467 \pm 110 $  & $  46.03   \pm 1.36    $   \\
080605  &   1.6398  &    $  3.4 \pm 0.28    $  & $  650 \pm 55  $  & $  45.61   \pm 0.93    $  &    090715B  &  3   &    $  1.09    \pm 0.17    $  & $  536 \pm 172 $  & $  46.93   \pm 1.66    $   \\
131105A &   1.686   &    $  4.75    \pm 0.16    $  & $  547.68  \pm 83.53   $  & $  44.94   \pm 1.09    $  &     080607 &   3.036   &    $  8.96    \pm 0.48    $  & $  1691    \pm 226 $  & $  46.85   \pm 1.04    $   \\
091020  &   1.71    &    $  0.11    \pm 0.034   $  & $  280 \pm 190 $  & $  47.75   \pm 3.13    $  &     081028 &   3.038   &    $  0.81    \pm 0.095   $  & $  234 \pm 93  $  & $  45.67   \pm 1.95    $   \\
100906  &   1.727   &    $  3.56    \pm 0.55    $  & $  387.23  \pm 244.07  $  & $  44.60   \pm 2.91    $  &     120922 &   3.1 &    $  1.59    \pm 0.18    $  & $  156.62  \pm 0.04    $  & $  44.19   \pm 0.86    $   \\
120119  &   1.728   &    $  4.62    \pm 0.59    $  & $  417.38  \pm 54.56   $  & $  44.46   \pm 1.04    $  &     020124 &   3.2 &    $  1.2 \pm 0.1 $  & $  448 \pm 148 $  & $  46.53   \pm 1.69    $   \\
110422  &   1.77    &    $  9.32    \pm 0.02    $  & $  421.04  \pm 13.85   $  & $  43.74   \pm 0.86    $  &     060526 &   3.21    &    $  0.12    \pm 0.06    $  & $  105 \pm 21  $  & $  46.25   \pm 1.34    $   \\
120326  &   1.798   &    $  0.44    \pm 0.02    $  & $  129.97  \pm 10.27   $  & $  44.81   \pm 0.92    $  &     080810 &   3.35    &    $  1.82    \pm 0.2 $  & $  1470    \pm 180 $  & $  48.40   \pm 1.02    $   \\
080514B &   1.8 &    $  2.027   \pm 0.48    $  & $  627 \pm 65  $  & $  46.17   \pm 1.00    $  &    110818   &  3.36    &    $  1.05    \pm 0.08    $  & $  1117.47     \pm 241.11  $  & $  48.47   \pm 1.28    $   \\
090902B &   1.822   &    $  32.38   \pm 1.01    $  & $  2187    \pm 31  $  & $  45.56   \pm 0.85    $  &     030323 &   3.37    &    $  0.12    \pm 0.04    $  & $  270 \pm 113 $  & $  48.10   \pm 2.07    $   \\
020127  &   1.9 &    $  0.38    \pm 0.01    $  & $  290 \pm 100 $  & $  46.55   \pm 1.74    $  &    971214   &  3.42    &    $  0.87    \pm 0.11    $  & $  685 \pm 133 $  & $  47.75   \pm 1.22    $   \\
080319C &   1.95    &    $  1.5 \pm 0.3 $  & $  906 \pm 272 $  & $  47.26   \pm 1.59    $  &    060707  &    3.425  &    $  0.23    \pm 0.04    $  & $  279 \pm 28  $  & $  47.47   \pm 0.98    $   \\
081008  &   1.9685  &    $  0.96    \pm 0.09    $  & $  261 \pm 52  $  & $  45.36   \pm 1.23    $  &     060115 &   3.53    &    $  0.25    \pm 0.04    $  & $  285 \pm 34  $  & $  47.45   \pm 1.02    $   \\
030226  &   1.98    &    $  1.3 \pm 0.1 $  & $  289 \pm 66  $  & $  45.23   \pm 1.32    $  &    090323  &    3.57   &    $  14.98   \pm 1.83    $  & $  1901    \pm 343 $  & $  46.65   \pm 1.17    $   \\
130612  &   2.006   &    $  0.08    \pm 0.01    $  & $  186.07  \pm 31.56   $  & $  47.43   \pm 1.14    $  &     100704 &   3.6 &    $  0.70    \pm 0.07    $  & $  809.60  \pm 135.70  $  & $  48.35   \pm 1.13    $   \\
000926  &   2.07    &    $  2.6 \pm 0.6 $  & $  310 \pm 20  $  & $  44.65   \pm 0.93    $  &    130514  &    3.6    &    $  1.88    \pm 0.25    $  & $  496.80  \pm 151.80  $  & $  46.34   \pm 1.60    $   \\
130610  &   2.092   &    $  0.82    \pm 0.05    $  & $  911.83  \pm 132.65  $  & $  47.98   \pm 1.07    $  &     130408 &   3.758   &    $  0.99    \pm 0.17    $  & $  1003.94     \pm 137.98  $  & $  48.42   \pm 1.06     $      \\
090926  &   2.1062  &    $  15.08   \pm 0.77    $  & $  974 \pm 50  $  & $  44.95   \pm 0.88    $  &     120802 &   3.796   &    $  0.43    \pm 0.07    $  & $  274.33  \pm 93.04   $  & $  46.85   \pm 1.73    $    \\
011211  &   2.14    &    $  0.5 \pm 0.06    $  & $  186 \pm 24  $  & $  45.48   \pm 1.03    $  &    100413   &  3.9 &    $  2.36    \pm 0.77    $  & $  1783.60     \pm 374.85  $  & $  48.61   \pm 1.31    $   \\
071020  &   2.145   &    $  0.87    \pm 0.4 $  & $  1013    \pm 160 $  & $  48.13   \pm 1.21    $  &     120909 &   3.93    &    $  2.69    \pm 0.23    $  & $  1651.55     \pm 123.25  $  & $  48.33   \pm 0.92     $      \\
050922C &   2.198   &    $  0.47    \pm 0.16    $  & $  415 \pm 111 $  & $  47.11   \pm 1.50    $  &     131117A    &   4.042   &    $  0.05    \pm 0.01    $  & $  221.85  \pm 37.31   $  & $  48.83   \pm 1.15    $    \\
121128  &   2.2 &    $  0.87    \pm 0.07    $  & $  243.20  \pm 12.80   $  & $  45.42   \pm 0.89    $  &     060206 &   4.048   &    $  0.14    \pm 0.03    $  & $  394 \pm 46  $  & $  48.82   \pm 1.02    $   \\
110205  &   2.22    &    $  3.32    \pm 0.68    $  & $  740.60  \pm 322.00  $  & $  46.10   \pm 2.11    $  &     090516 &   4.109   &    $  1.96    \pm 0.38    $  & $  971 \pm 390 $  & $  47.70   \pm 1.98    $   \\
130505  &   2.27    &    $  4.56    \pm 0.09    $  & $  2063.37     \pm 101.37  $  & $  47.74   \pm 0.88     $  &   080916C &   4.35    &    $  10.13   \pm 2.13    $  & $  2646    \pm 566 $  & $  47.88   \pm 1.29     $      \\
060124  &   2.296   &    $  3.4 \pm 0.5 $  & $  784 \pm 285 $  & $  46.21   \pm 1.82    $  &    000131  &    4.5    &    $  4.7 \pm 0.8 $  & $  987 \pm 416 $  & $  46.86   \pm 2.05    $   \\
021004  &   2.3 &    $  0.27    \pm 0.04    $  & $  266 \pm 117 $  & $  46.89   \pm 2.13    $  &    111008   &  5   &    $  1.06    \pm 0.11    $  & $  894.00  \pm 240.00  $  & $  48.38   \pm 1.46    $   \\
051109A &   2.346   &    $  0.51    \pm 0.05    $  & $  539 \pm 200 $  & $  47.57   \pm 1.85    $  &     060927 &   5.6 &    $  0.27    \pm 0.04    $  & $  475 \pm 47  $  & $  48.76   \pm 0.97    $   \\
060908  &   2.43    &    $  0.73    \pm 0.07    $  & $  514 \pm 102 $  & $  47.12   \pm 1.23    $  &     130606 &   5.91    &    $  0.49    \pm 0.09    $  & $  2031.54     \pm 483.70  $  & $  50.94   \pm 1.37     $      \\
080413  &   2.433   &    $  0.56    \pm 0.14    $  & $  584 \pm 180 $  & $  47.65   \pm 1.63    $  &     050904 &   6.29    &    $  2   \pm 0.2 $  & $  3178    \pm 1094    $  & $  50.33   \pm 1.75    $   \\
090812  &   2.452   &    $  3.077   \pm 0.53    $  & $  2000    \pm 700 $  & $  48.17   \pm 1.77    $  &     080913 &   6.695   &    $  0.12    \pm 0.035   $  & $  710 \pm 350 $  & $  50.57   \pm 2.36    $   \\
100728B &   2.453   &    $  0.29    \pm 0.01    $  & $  359.11  \pm 48.34   $  & $  47.44   \pm 1.04    $  &     090423 &   8.2 &    $  0.12    \pm 0.032   $  & $  491 \pm 200 $  & $  50.05   \pm 2.01    $   \\
\hline \caption{\label{calibratedGRB}The 90 calibrated GRBs with
redshift, bolometric fluence, peak energy in cosmological rest frame
and distance moduli. The 1\,$\sigma$ uncertainties are also given.}
\centering
\end{longtable}
}


\begin{thebibliography}{}
\bibitem[\protect\citeauthoryear{Amanullah et al.}{2010}]{Amanullah2010}
Amanullah, R., et al., 2010, ApJ, 716, 712

\bibitem[\protect\citeauthoryear{Amati, Frontera \& Guidorzi}{2009}]{Amati2009} Amati L.,
Frontera F., Guidorzi C., 2009, A\&A, 508, 173


\bibitem[\protect\citeauthoryear{Amati et al.}{2002}]{Amati2002} Amati L., et al., 2002, A\&A, 390, 81


\bibitem[\protect\citeauthoryear{Amati et al.}{2008}]{Amati2008}
Amati L., Guidorzi C., Frontera F., Della Valle M., Finelli F., Landi R.,
Montanari E., 2008, MNRAS, 391, 577


\bibitem[\protect\citeauthoryear{Anderson et al.}{2014}]{Anderson2014}
Anderson L., et al., 2014, MNRAS, 441, 24


\bibitem[\protect\citeauthoryear{Band et al.}{1993}]{Band1993}
Band D., et al., 1993, ApJ, 413, 281


\bibitem[\protect\citeauthoryear{Barkana \& Loeb}{2004}]{Barkana2004}
Barkana R., Loeb A., 2004, ApJ, 601, 64


\bibitem[Basa et al.(2008)]{Basa2008} Basa, S.,
Wei, J., Paul, J., Zhang, S.~N., \& Svom Collaboration 2008, SF2A-2008, 161


\bibitem[\protect\citeauthoryear{Basilakos \& Perivolaropoulos}{2008}]{Basilakos08}Basilakos S., Perivolaropoulos L., 2008, MNRAS, 391, 411


\bibitem[\protect\citeauthoryear{Beutler et al.}{2011}]{Beutler2011}
Beutler F., et al., 2011, MNRAS, 416, 3017


\bibitem[\protect\citeauthoryear{Bloom, Frail \& Sari}{2001}]{Bloom2001}
Bloom J.~S., Frail D.~A., Sari R., 2001, AJ, 121, 2879

\bibitem[\protect\citeauthoryear{Cardone et al.}{2010}]{Cardone2010}
Cardone V.~F., Dainotti M.~G., Capozziello S., Willingale R., 2010, MNRAS, 408, 1181


\bibitem[\protect\citeauthoryear{Ciardi \& Loeb}{2000}]{Ciardi2000}
Ciardi B., Loeb A., 2000, ApJ, 540, 687


\bibitem[\protect\citeauthoryear{Cole et al.}{2005}]{Cole2005}
Cole S., et al., 2005, MNRAS, 362, 505

\bibitem[\protect\citeauthoryear{Collazzi}{2012}]{Collazzi2012}
Collazzi C., et al., 2012, GRB Coordinates Network Circular, 13145, 1


\bibitem[\protect\citeauthoryear{Conley et al.}{2011}]{Conley2011}
Conley A., et al., 2011, ApJS, 192, 1


\bibitem[\protect\citeauthoryear{Cucchiara et al.}{2011}]{Cucchiara2011}
Cucchiara A., et al., 2011q, ApJ, 743, 154


\bibitem[\protect\citeauthoryear{D'Agostini}{2005}]{D'Agostini2005}
D'Agostini G., 2005, physics.., arXiv:physics/0511182


\bibitem[\protect\citeauthoryear{Dai, Liang \& Xu}{2004}]{Dai2004}
Dai Z.~G., Liang E.~W., Xu D., 2004, ApJ, 612, L101


\bibitem[\protect\citeauthoryear{Dainotti, Cardone \& Capozziello}{2008}]{Dainotti2008}
Dainotti M.~G., Cardone V.~F., Capozziello S., 2008, MNRAS, 391, L79


\bibitem[\protect\citeauthoryear{Dainotti et al.}{2010}]{Dainotti2010}
Dainotti M.~G., Willingale R., Capozziello
S., Fabrizio Cardone V., Ostrowski M., 2010, ApJ, 722, L215


\bibitem[\protect\citeauthoryear{Eisenstein et al.}{2005}]{Eisenstein2005}
Eisenstein D.~J., et al., 2005, ApJ, 633, 560


\bibitem[\protect\citeauthoryear{Fitzpatrick et al.}{2013a}]{Fitzpatrick2013a}
Fitzpatrick G., et al., 2013a, GRB Coordinates Network Circular,
14858, 1

\bibitem[\protect\citeauthoryear{Fitzpatrick et al.}{2013b}]{Fitzpatrick2013b}
Fitzpatrick G., et al., 2013b, GRB Coordinates Network Circular,
14896, 1

\bibitem[\protect\citeauthoryear{Fitzpatrick et al.}{2013c}]{Fitzpatrick2013c}
Fitzpatrick G., et al., 2013c, GRB Coordinates Network Circular,
15455, 1


\bibitem[\protect\citeauthoryear{Friedman \& Bloom}{2005}]{Friedman2005}
Friedman A.~S., Bloom J.~S., 2005, ApJ, 627, 1


\bibitem[\protect\citeauthoryear{Gallerani et al.}{2008}]{Gallerani2008}
Gallerani S., Salvaterra R., Ferrara A., Choudhury T.~R., 2008, MNRAS, 388, L84

\bibitem[\protect\citeauthoryear{Gehrels et al.}{2009}]{Gehrels09} Gehrels N., Ramirez-Ruiz E., Fox D. B., 2009, ARA\&A, 47, 567

\bibitem[\protect\citeauthoryear{Gendre et al.}{2013}]{Gendre2013}
Gendre B., et al., 2013, ApJ, 766, 30


\bibitem[\protect\citeauthoryear{Ghirlanda et al.}{2004a}]{Ghirlanda2004a}
Ghirlanda G., Ghisellini G., Lazzati D., 2004, ApJ, 616, 331


\bibitem[\protect\citeauthoryear{Ghirlanda et al.}{2004b}]{Ghirlanda2004b}
Ghirlanda G., Ghisellini G., Lazzati D., Firmani C., 2004, ApJ, 613, L13

\bibitem[\protect\citeauthoryear{Ghirlanda}{2009}]{Ghirlanda2009}
Ghirlanda G., 2009, AIPC, 1111, 579

\bibitem[\protect\citeauthoryear{Golenetskii}{2010a}]{Golenetskii2010a}
Golenetskii S., et al., 2010a, GRB Coordinates Network Circular,
10882, 1

\bibitem[\protect\citeauthoryear{Golenetskii}{2010b}]{Golenetskii2010b}
Golenetskii S., et al., 2010b, GRB Coordinates Network Circular,
10937, 1

\bibitem[\protect\citeauthoryear{Golenetskii}{2010c}]{Golenetskii2010c}
Golenetskii S., et al., 2010c, GRB Coordinates Network Circular,
11119, 1

\bibitem[\protect\citeauthoryear{Golenetskii}{2010d}]{Golenetskii2010d}
Golenetskii S., et al., 2010d, GRB Coordinates Network Circular,
11251, 1

\bibitem[\protect\citeauthoryear{Golenetskii}{2011a}]{Golenetskii2011a}
Golenetskii S., et al., 2011a, GRB Coordinates Network Circular,
11723, 1

\bibitem[\protect\citeauthoryear{Golenetskii}{2011b}]{Golenetskii2011b}
Golenetskii S., et al., 2011b, GRB Coordinates Network Circular,
11971, 1

\bibitem[\protect\citeauthoryear{Golenetskii}{2011c}]{Golenetskii2011c}
Golenetskii S., et al., 2011c, GRB Coordinates Network Circular,
12008, 1

\bibitem[\protect\citeauthoryear{Golenetskii}{2011d}]{Golenetskii2011d}
Golenetskii S., et al., 2011d, GRB Coordinates Network Circular,
12166, 1

\bibitem[\protect\citeauthoryear{Golenetskii}{2011e}]{Golenetskii2011e}
Golenetskii S., et al., 2011e, GRB Coordinates Network Circular,
12223, 1

\bibitem[\protect\citeauthoryear{Golenetskii}{2011f}]{Golenetskii2011f}
Golenetskii S., et al., 2011f, GRB Coordinates Network Circular,
12433, 1

\bibitem[\protect\citeauthoryear{Golenetskii}{2011g}]{Golenetskii2011g}
Golenetskii S., et al., 2011g, GRB Coordinates Network Circular,
12872, 1

\bibitem[\protect\citeauthoryear{Golenetskii}{2012a}]{Golenetskii2012a}
Golenetskii S., et al., 2012a, GRB Coordinates Network Circular,
13736, 1

\bibitem[\protect\citeauthoryear{Golenetskii}{2012b}]{Golenetskii2012b}
Golenetskii S., et al., 2012b, GRB Coordinates Network Circular,
14010, 1

\bibitem[\protect\citeauthoryear{Golenetskii}{2013a}]{Golenetskii2013a}
Golenetskii S., et al., 2013a, GRB Coordinates Network Circular,
14368, 1

\bibitem[\protect\citeauthoryear{Golenetskii}{2013b}]{Golenetskii2013b}
Golenetskii S., et al., 2013b, GRB Coordinates Network Circular,
14487, 1

\bibitem[\protect\citeauthoryear{Golenetskii}{2013c}]{Golenetskii2013c}
Golenetskii S., et al., 2013c, GRB Coordinates Network Circular,
14575, 1

\bibitem[\protect\citeauthoryear{Golenetskii}{2013d}]{Golenetskii2013e}
Golenetskii S., et al., 2013d, GRB Coordinates Network Circular,
14808, 1

\bibitem[\protect\citeauthoryear{Golenetskii}{2013e}]{Golenetskii2013f}
Golenetskii S., et al., 2013e, GRB Coordinates Network Circular,
14958, 1

\bibitem[\protect\citeauthoryear{Golenetskii}{2013f}]{Golenetskii2013g}
Golenetskii S., et al., 2013f, GRB Coordinates Network Circular,
15145, 1

\bibitem[\protect\citeauthoryear{Golenetskii}{2013g}]{Golenetskii2013h}
Golenetskii S., et al., 2013g, GRB Coordinates Network Circular,
15203, 1

\bibitem[\protect\citeauthoryear{Golenetskii}{2013h}]{Golenetskii2013i}
Golenetskii S., et al., 2013h, GRB Coordinates Network Circular,
15413, 1


\bibitem[G{\"o}tz et al.(2009)]{Gotz2009} G{\"o}tz, D., Paul,
J., Basa, S., et al.\ 2009, American Institute of Physics Conference
Series, 1133, 25



\bibitem[\protect\citeauthoryear{Isobe et al.}{1990}]{Isobe1990}
Isobe T., Feigelson E.~D., Akritas M.~G., Babu G.~J., 1990, ApJ, 364, 104


\bibitem[\protect\citeauthoryear{Kazin et al.}{2014}]{Kazin2014}
Kazin E.~A., et al., 2014, MNRAS, 441, 3524

\bibitem[\protect\citeauthoryear{Kodama et al.}{2008}]{Kodama2008}
Kodama Y., Yonetoku D., Murakami T., Tanabe S., Tsutsui R., Nakamura
T., 2008, MNRAS, 391, L1

\bibitem[\protect\citeauthoryear{Komatsu et al.}{2011}]{Komatsu11}
Komatsu E., Smith K. M., Dunkley J., et al., 2011, ApJS, 192, 18

\bibitem[\protect\citeauthoryear{Kowalski et al.}{2008}]{Kowalski2008}
Kowalski, M., et al., 2008, ApJ, 686, 749

\bibitem[\protect\citeauthoryear{Krimm et al.}{2012a}]{Krimm2012a}
Krimm H. A., et al., 2012a, GRB Coordinates Network Circular, 13517,
1

\bibitem[\protect\citeauthoryear{Krimm et al.}{2012b}]{Krimm2012b}
Krimm H. A., et al., 2012b, GRB Coordinates Network Circular, 13634,
1

\bibitem[\protect\citeauthoryear{Krimm et al.}{2012c}]{Krimm2012c}
Krimm H. A., et al., 2012c, GRB Coordinates Network Circular, 13806,
1

\bibitem[\protect\citeauthoryear{Krimm et al.}{2013}]{Krimm2013}
Krimm H. A., et al., 2013, GRB Coordinates Network Circular, 15499,
1


\bibitem[\protect\citeauthoryear{Lamb \& Reichart}{2000}]{Lamb2000}
Lamb D.~Q., Reichart D.~E., 2000, ApJ, 536, 1



\bibitem[\protect\citeauthoryear{Li \& Lin}{2015}]{Li2015}
Li M.-H., Lin H.-N., 2015, ApJ, 807, 76


\bibitem[\protect\citeauthoryear{Liang \& Zhang}{2005}]{Liang2005}
Liang E., Zhang B., 2005, ApJ, 633, 611


\bibitem[\protect\citeauthoryear{Liang et al.}{2008}]{Liang2008}
Liang N., Xiao W.~K., Liu Y., Zhang S.~N., 2008, ApJ, 685, 354


\bibitem[\protect\citeauthoryear{Lin, Li, \& Change}{2015}]{Lin2015}
Lin H.-N., Li X., Change Z., 2015, arXiv:1507.06662


\bibitem[\protect\citeauthoryear{M\'{e}sz\'{a}ros}{2006}]{Meszaros06}
M\'{e}sz\'{a}ros P., 2006, Rep. Prog. Phys., 69, 2259


\bibitem[\protect\citeauthoryear{Palshin et al.}{2013}]{Palshin2013}
Palshin V., et al., 2013, GRB Coordinates Network Circular, 14702, 1


\bibitem[Paul et al.(2011)]{Paul2011} Paul, J., Wei, J., Basa,
S., \& Zhang, S.-N.\ 2011, Comptes Rendus Physique, 12, 298


\bibitem[\protect\citeauthoryear{Pelassa et al.}{2011}]{Pelassa2011}
Pelassa V., et al., 2011, GRB Coordinates Network Circular, 12545, 1

\bibitem[\protect\citeauthoryear{Perlmutter et al.}{1999}]{Perlmutter1999}
Perlmutter S., et al., 1999, ApJ, 517, 565

\bibitem[\protect\citeauthoryear{Pescalli et al.}{2015}]{Pescalli2015}
Pescalli A., et al., 2015, arXiv: 1506.05463v1

\bibitem[\protect\citeauthoryear{Planck Collaboration}{2013}]{Planck13}
Planck Collaboration,
Ade, P.~A.~R., Aghanim, N., et al.\ 2014, \aap, 571, A16



\bibitem[\protect\citeauthoryear{Planck Collaboration}{2015}]{Planck15}
Planck Collaboration, et al., 2015, arXiv:1502.01589



\bibitem[\protect\citeauthoryear{Porciani \& Madau}{2001}]{Porciani2001}
Porciani C., Madau P., 2001, ApJ, 548, 522


\bibitem[\protect\citeauthoryear{Qi \& Lu}{2010}]{Qi2010}
Qi S., Lu T., 2010, ApJ, 717, 1274


\bibitem[\protect\citeauthoryear{Qi, Lu \& Wang}{2009}]{Qi2009}
Qi S., Lu T., Wang F.-Y., 2009, MNRAS, 398, L78


\bibitem[\protect\citeauthoryear{Riess et al.}{1998}]{Riess1998}
Riess A.~G., et al., 1998, AJ, 116, 1009


\bibitem[\protect\citeauthoryear{Ross et al.}{2015}]{Ross2015}
Ross A.~J., Samushia L., Howlett C., Percival W.~J., Burden A., Manera M.,
2015, MNRAS, 449, 835


\bibitem[\protect\citeauthoryear{Salvaterra et al.}{2009}]{Salvaterra2009}
Salvaterra R., et al., 2009, Natur, 461, 1258

\bibitem[\protect\citeauthoryear{Salvaterra et al.}{2012}]{Salvaterra2012}
Salvaterra R., et al., 2012, ApJ, 749, 68

\bibitem[\protect\citeauthoryear{Schaefer}{2007}]{Schaefer2007}
Schaefer B.~E., 2007, ApJ, 660, 16


\bibitem[\protect\citeauthoryear{Schaefer}{2003}]{Schaefer2003}
Schaefer B.~E., 2003, ApJ, 583, L67

\bibitem[\protect\citeauthoryear{Stamatikos et al.}{2012}]{Stamatikos2012}
Stamatikos M., et al., 2012, GRB Coordinates Network Circular,
13559, 1

\bibitem[\protect\citeauthoryear{Sugita et al.}{2010}]{Sugita2010}
Sugita S., et al., 2010, GRB Coordinates Network Circular, 10604, 1

\bibitem[\protect\citeauthoryear{Suzuki et al.}{2012}]{Suzuki2012}
Suzuki N., et al., 2012, ApJ, 746, 85

\bibitem[Tanvir et al.(2009)]{Tanvir2009} Tanvir, N.~R., Fox,
D.~B., Levan, A.~J., et al., 2009, Natur, 461, 1254

\bibitem[\protect\citeauthoryear{Totani}{1997}]{Totani1997}
Totani T., 1997, ApJ, 486, L71

\bibitem[\protect\citeauthoryear{Totani et al.}{2006}]{Totani2006}
Totani T., Kawai N., Kosugi G., Aoki K., Yamada T., Iye M., Ohta K.,
Hattori T., 2006, PASJ, 58, 485

\bibitem[\protect\citeauthoryear{Kienlin et al.}{2010}]{Kienlin2010}
von Kienlin A., et al., 2010, GRB Coordinates Network Circular,
11015, 1

\bibitem[\protect\citeauthoryear{Kienlin et al.}{2013}]{Kienlin2013}
von Kienlin A., et al., 2013, GRB Coordinates Network Circular,
14473, 1

\bibitem[\protect\citeauthoryear{Kienlin et al.}{2014}]{Kienlin2014}
von Kienlin A., et al., 2014, GRB Coordinates Network Circular,
15796, 1


\bibitem[\protect\citeauthoryear{Wang}{2013}]{Wang2013}
Wang F.~Y., 2013, A\&A, 556, A90


\bibitem[\protect\citeauthoryear{Wang et al.}{2012}]{Wang2012}
Wang F.~Y., Bromm V., Greif T.~H., Stacy A., Dai Z.~G., Loeb A., Cheng
K.~S., 2012, ApJ, 760, 27


\bibitem[\protect\citeauthoryear{Wang \& Dai}{2009}]{Wang2009}
Wang F.~Y., Dai Z.~G., 2009, MNRAS, 400, L10


\bibitem[\protect\citeauthoryear{Wang, Dai \& Zhu}{2007}]{Wang2007}
Wang F.~Y., Dai Z.~G., Zhu Z.-H., 2007, ApJ, 667, 1

\bibitem[\protect\citeauthoryear{Wang et al.}{2015}]{Wang14}
Wang F.~Y., Dai Z.~G., Liang E.~W., 2015, NewAR, 67, 1


\bibitem[\protect\citeauthoryear{Wang \& Dai}{2011a}]{Wang2011a}
Wang F.~Y., Dai Z.~G., 2011a, ApJ, 727, L34


\bibitem[\protect\citeauthoryear{Wang}{2012}]{Wang12}
Wang F.~Y., 2012, A\&A, 543, A91


\bibitem[\protect\citeauthoryear{Wang \& Dai}{2014}]{WangD2014}
Wang F.~Y., Dai Z.~G., 2014, PhRvD, 89, 023004


\bibitem[\protect\citeauthoryear{Wang, Qi \& Dai}{2011b}]{Wang2011b}
Wang F.-Y., Qi S., Dai Z.-G., 2011, MNRAS, 415, 3423


\bibitem[\protect\citeauthoryear{Wang \& Dai}{2011c}]{Wang11}
Wang F.~Y., Dai Z.~G., 2011c, A\&A, 536, A96


\bibitem[\protect\citeauthoryear{Wang \& Wang}{2014a}]{Wang2014a}
Wang, J.~S., \& Wang, F.~Y.\ 2014, \mnras, 443, 1680


\bibitem[\protect\citeauthoryear{Wang \& Wang}{2014b}]{Wang2014b}
Wang J.~S., Wang F.~Y., 2014, A\&A, 564, A137


\bibitem[\protect\citeauthoryear{Wei \& Gao}{2003}]{Wei2003}
Wei D.~M., Gao W.~H., 2003, MNRAS, 345, 743


\bibitem[\protect\citeauthoryear{Wei}{2010}]{Wei2010} Wei H.,
2010, JCAP, 8, 20


\bibitem[\protect\citeauthoryear{Wijers et al.}{1998}]{Wijers1998}
Wijers R.~A.~M.~J., Bloom J.~S., Bagla J.~S., Natarajan P., 1998, MNRAS,
294, L13

\bibitem[\protect\citeauthoryear{Xiong et al.}{2011}]{Xiong2011}
Xiong S., et al., 2011, GRB Coordinates Network Circular, 12287, 1

\bibitem[\protect\citeauthoryear{Xiong et al.}{2013a}]{Xiong2013a}
Xiong S., et al., 2013a, GRB Coordinates Network Circular, 14429, 1

\bibitem[\protect\citeauthoryear{Xiong et al.}{2013b}]{Xiong2013b}
Xiong S., et al. 2013b, GRB Coordinates Network Circular, 14674, 1

\bibitem[\protect\citeauthoryear{Younes et al.}{2013}]{Younes2013}
Younes G., et al., 2013, GRB Coordinates Network Circular, 14219, 1

\bibitem[\protect\citeauthoryear{Zhang}{2007}]{Zhang07}
Zhang B., 2007, Chin.~J.~Astron.~Astrophys., 7, 1


\bibitem[\protect\citeauthoryear{Zhang et al.}{2014}]{Zhang2014}
Zhang B. B., 2014, GRB Coordinates Network Circular, 15833, 1

\end{thebibliography}
\end{document}